\documentclass[reprint,prl,aps,superscriptaddress]{revtex4-1}
\usepackage{times}
\usepackage{graphicx}
\usepackage{amsmath, braket, amsfonts}
\usepackage{amssymb}
\usepackage{sidecap}
\usepackage{bm, color, ulem}
\usepackage{xcolor}

\DeclareUnicodeCharacter{03BD}{{$\nu$}}
\begin{document}
\title{Isospin magnetism and spin-polarized superconductivity in Bernal bilayer graphene}
\author{Haoxin Zhou}   
\email{Present address: Department of Applied Physics and Material Science, California Institute of Technology, Pasadena, CA 91125}
\affiliation{Department of Physics, University of California at Santa Barbara, Santa Barbara CA 93106, USA}
\author{Ludwig Holleis} \affiliation{Department of Physics, University of California at Santa Barbara, Santa Barbara CA 93106, USA}
\author{Yu Saito}   
\affiliation{Department of Physics, University of California at Santa Barbara, Santa Barbara CA 93106, USA}
\author{Liam Cohen}   
\affiliation{Department of Physics, University of California at Santa Barbara, Santa Barbara CA 93106, USA}
\author{William Huynh}   
\affiliation{Department of Physics, University of California at Santa Barbara, Santa Barbara CA 93106, USA}
\author{Caitlin L. Patterson}   
\affiliation{Department of Physics, University of California at Santa Barbara, Santa Barbara CA 93106, USA}
\author{Fangyuan Yang}   
\affiliation{Department of Physics, University of California at Santa Barbara, Santa Barbara CA 93106, USA}
\author{Takashi Taniguchi}
\affiliation{International Center for Materials Nanoarchitectonics,
National Institute for Materials Science,  1-1 Namiki, Tsukuba 305-0044, Japan}
\author{Kenji Watanabe}
\affiliation{Research Center for Functional Materials,
National Institute for Materials Science, 1-1 Namiki, Tsukuba 305-0044, Japan}
\author{Andrea F. Young}
\email{andrea@physics.ucsb.edu}
\affiliation{Department of Physics, University of California at Santa Barbara, Santa Barbara CA 93106, USA}

\date{\today}

\begin{abstract}
We report the observation of spin-polarized superconductivity in Bernal bilayer graphene when doped to a saddle-point van Hove singularity generated by large applied perpendicular electric field.   We observe a cascade of electrostatic gate-tuned transitions between electronic phases distinguished by their polarization within the isospin space defined by the combination of the spin and momentum-space valley degrees of freedom. 
While all of these phases are metallic at zero magnetic field, we observe a transition to a superconducting state at finite $B_\parallel\approx 150$mT applied parallel to the two dimensional sheet. 
Superconductivity occurs near a symmetry breaking transition, and exists exclusively above the $B_\parallel$-limit expected of a paramagnetic superconductor with the observed $T_C\approx 30$mK, implying a spin-triplet order parameter. 
\end{abstract}
\maketitle

Spin-triplet superconductors are rare in nature.  This scarcity is traceable, at least in part, to the inapplicability of Anderson's theorem\cite{anderson_theory_1959} that renders conventional superconductors immune to disorder. Realizing spin triplet superconductivity thus places stringent bounds on materials quality. 
Recently, graphene-based two dimensional materials have emerged as a novel platform for superconductivity\cite{cao_unconventional_2018,yankowitz_dynamic_2018,hao_electric_2021,park_tunable_2021,zhou_superconductivity_2021}.  
In particular, two varieties of graphene trilayer---one rotationally faulted\cite{cao_pauli-limit_2021}, and one in a metastable rhombohedral stacking order\cite{zhou_superconductivity_2021}---have shown superconducting states that are resilient to magnetic fields. Both appear to persist above the nominal critical magnetic field\cite{clogston_upper_1962,chandrasekhar_note_1962} for a paramagnetic superconductor derived by comparing their critical temperature to the Zeeman energy, suggestive of a spin triplet order parameter. 
Unfortunately, neither of these materials represents a structural ground state.
Rotationally faulted structures are generally unstable, limiting sample uniformity and, consequently, reproducibility\cite{balents_superconductivity_2020}. Rhombohedral stacking orders, meanwhile, are only metastable, allowing uniform structures to be produced but at great cost in practical yield of working devices.  These drawbacks hamper efforts to systematically vary experimental parameters, and to build more complex devices making use of the array of gate tuned phases available in these materials. 

\begin{figure*}
    \centering
    \includegraphics{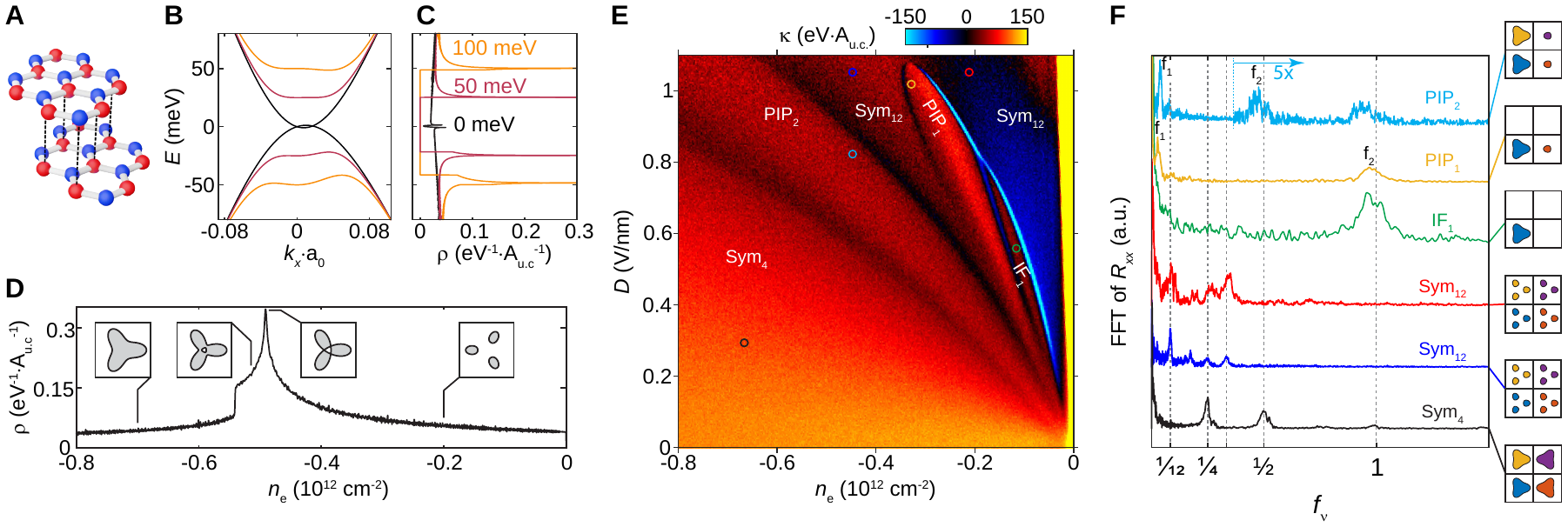}
    \caption{\textbf{Broken isospin symmetry phases in Bernal bilayer graphene.} 
    \textbf{A}, Lattice structure of Bernal bilayer graphene. \textbf{B}, Band structure calculated within a tight binding model\cite{jung_accurate_2014} near the Brillouin zone corner.  $a_0=2.46\AA$ is the graphene lattice constant. 
    \textbf{C}, Noninteracting density of states, $\rho$.  Different curves are calculated for varying interlayer potential difference as marked.  $A_{u.c.}\approx5.24 \AA^2$ is the graphene unit cell area.
    \textbf{D}, $\rho$ as a function of carrier density calculated\cite{jung_accurate_2014} for fixed interlayer potential difference of 50mV. Insets: Fermi contours at the indicated values of $n_{\rm e}$, plotted on the interval $(-0.06,0.06) a_0^{-1}$ for both $k_x$ and $k_y$. 
    \textbf{E, } Experimentally determined inverse compressibility at zero magnetic field.
    \textbf{F, } Fourier transform of $R_{xx}(1/B_\perp)$ measured at the ($n_{\rm e}$,$D$) points indicated by the colored symbols in panel E.  
    The x-axis is the frequency normalized to $n_{\rm e}$. 
    Peaks reflect fractional area of the Fermi sea enclosed by a phase-coherent orbit. 
    For each trace, we show a schematic of the spin-dependent Fermi surfaces in the two valleys inspired by a rigid-band Stoner model of the type considered in \cite{zhou_half-_2021}.}
    \label{fig:1}
\end{figure*}

Here, we report magnetic field-induced superconductivity in Bernal bilayer graphene (BBG), whose crystal structure is shown in Fig. 1A. 
Bilayer graphene has been the subject of hundreds of experimental studies since its original experimental description in 2006\cite{novoselov_unconventional_2006}. However, prior explorations of electron correlation physics have focused on instabilities of the parabolic band touching that occurs in the absence of an applied displacement field\cite{weitz_broken-symmetry_2010,mayorov_interaction-driven_2011,bao_evidence_2012,freitag_spontaneously_2012}.  
When a perpendicular electric displacement field ($D$) is applied, the parabolic band touching is replaced by a band gap (Fig. \ref{fig:1}B), with van Hove singularities characterized by divergent single particle density of states appearing near the band edge.  Energy bands and associated single particle density of states calculated within a four-band tight binding model\cite{jung_accurate_2014} are plotted in Figs. \ref{fig:1}B-C.  
Fig. \ref{fig:1}D shows the calculated density of states and select Fermi contours at interlayer potential difference of 50 meV, corresponding\cite{zhang_direct_2009} to $D\approx 0.5$V/nm. 
A van Hove singularity occurs at a carrier density of $n_{\rm e}\approx -0.5 \times 10^{12} $cm$^{-2}$, where three low-density Fermi pockets merge into an annulus. We note that our choice of tight binding parameters, derived from numerical band structure modeling\cite{jung_accurate_2014}, has not been quantitatively bench marked to experiment in the regime of interest. However, the existence of a saddle point van Hove singularity in this approximate density regime is expected to be generic.  

The electronic structure of BBG resembles that of rhombohedral graphite multilayers\cite{koshino_trigonal_2009,zhang_band_2010}.  However, BBG is considerably easier to manufacture due to its structural stability.  
Our devices consist of a BBG channel encapsulated in single crystal hexagonal boron nitride gate dielectrics in which the charge carrier density $n_{\rm e}$ and electrical displacement field $D$ are controlled by single crystal graphite top and bottom gates\cite{zibrov_tunable_2017}. 
We report data from two devices which show nearly identical behavior.  Data shown in the main text are from Device A, with data from device B is shown in \cite{zhou_supplementary_2021}.  

Fig. \ref{fig:1}E shows inverse electronic compressibility $\kappa=\partial \mu/\partial n_{\rm e}$\cite{eisenstein_compressibility_1994,zibrov_tunable_2017} measured for small hole doping.
A series of transitions are visible as dips in the inverse compressibility, accompanied by concomitant sharp changes in the electrical resistivity (see Fig. \ref{fig:s:r_cp_linecuts} in \cite{zhou_supplementary_2021} for additional data).  
High resolution quantum oscillation data show that these features are associated with changes in the Fermi surface topology linked to breaking of the spin- and valley symmetries. 
Fig. 1F shows the Fourier transform of the magnetoresistance ( see Fig. \ref{fig:s:original_osc} in \cite{zhou_supplementary_2021} for additional data), $R(1/B_\perp)$, measured at different $(n_{\rm e},D)$ points indicated in panel 1E. 
Fourier transforms are plotted as a function of the oscillation frequency normalized to the total carrier density, which we denote $f_\nu$.  $f_\nu$ corresponds to the fraction of the Luttinger volume encircled by the phase coherent orbit that generates a given oscillation peak. To determine $n_{\rm e}$, we use the geometric capacitances of the gates as well as the spectroscopically determined\cite{zhang_direct_2009} band gap $\Delta$, so that for hole doping $n_{\rm e}=c_{\rm t} v_{\rm t}+c_{\rm b} v_{\rm b} +\Delta/2$. 
At high $|n_{\rm e}|$, a prominent peak is visible at $f_\nu=0.25$, along with associated harmonics.
This is consistent with a state preserving the four-fold combined spin- and valley degeneracy of the honeycomb lattice which has a simple Fermi surface in each of the four isospin flavors, as shown schematically at bottom right of Fig. \ref{fig:1}F.  We denote this symmetric phase Sym$_4$. At low densities and high $D$, in contrast, the strongest peak occurs at $f_\nu=1/12$.  
This is again consistent with intact isospin symmetry, but in the regime of density where trigonal warping produces three Fermi pockets within a single isospin flavor. This phase is denoted Sym$_{12}$.  In the single particle picture, Sym$_{12}$ and Sym$_4$ are the phases that straddle the van Hove singularity (Fig. \ref{fig:1}D).

We also identify regions with lowered degeneracy.  The first, which appears at low density, is characterized by a broad peak at $f_\nu=1$ and corresponds to a quarter metal with a single, fully isospin polarized Fermi surface (we denote this isospin ferromagnet IF$_1$).  
Adjoining IF$_1$ is a phase with a strong peak at $f_\nu$ slightly smaller than 1.  
We associate this signature with a partially isospin polarized phase featuring a large Fermi surface of one isospin flavor and one (or possibly more) smaller Fermi surfaces in a second, and denote it PIP$_1$.  
An additional phase appears as a sash at densities intermediate between Sym$_4$ and Sym$_{12}$. 
This phase shows two prominent peaks at $f_\nu=f_1$ and at $f_\nu=f_2$ such that $f_1+f_2=0.5$. 
We dub this phase PIP$_2$ and associate it with the existence of a single Fermi surface in each of two majority and two minority isospin flavors.  Possible fermi surface topologies for the observed phases are depicted in Fig. \ref{fig:1}F, for the case where spin and valley remain good quantum numbers.  Notably, these schematic depictions do not allow for the possibility of inter-valley coherence, which is theoretically possible but cannot be unambiguously determined from the quantum oscillation data alone.  

\begin{figure*}
    \centering
    \includegraphics{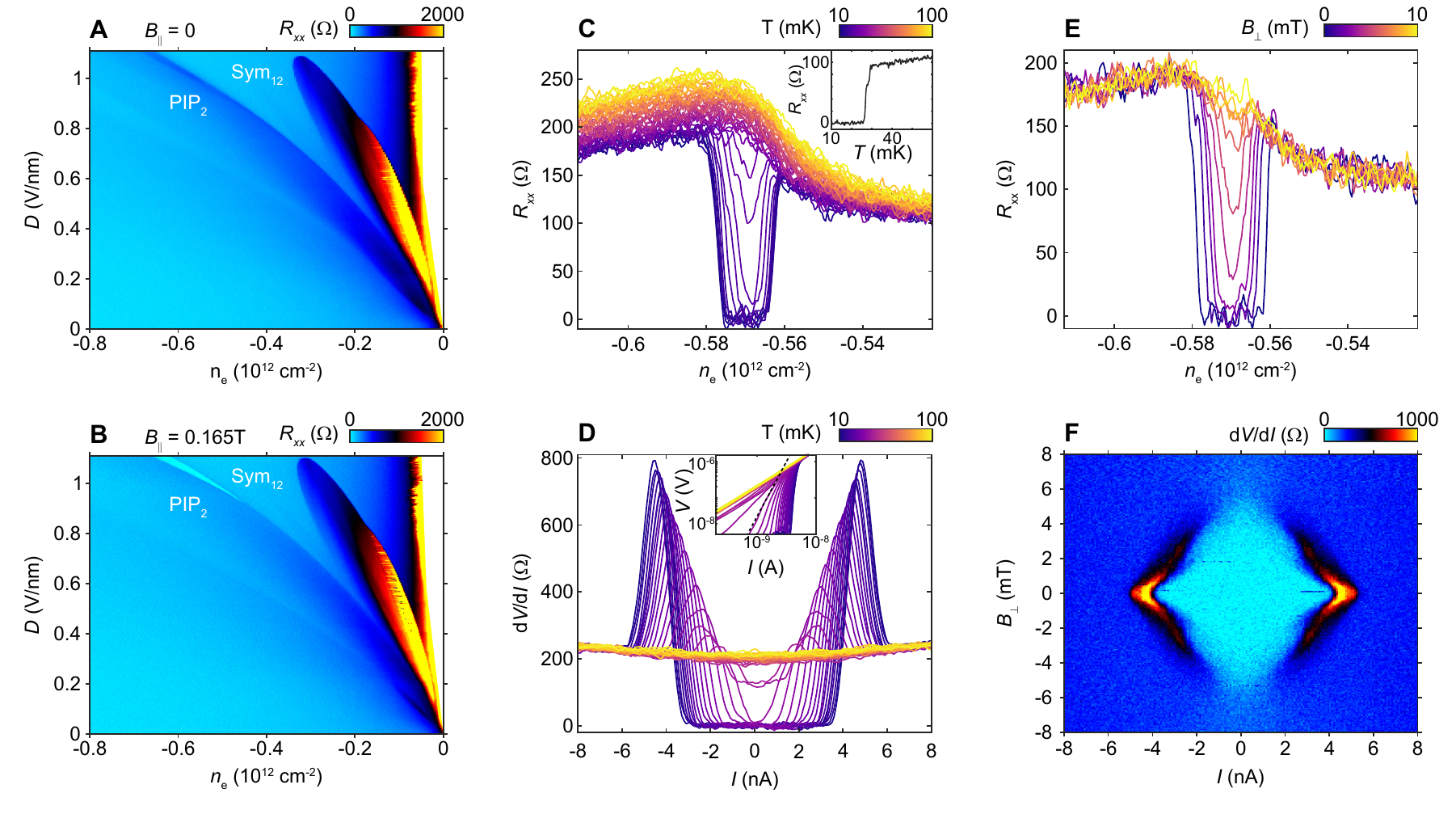}
    \caption{\textbf{Magnetic-field induced superconductivity.} 
    \textbf{A}, $R_{xx}$ measured at fixed $B_\parallel=0$ at a nominal temperature of 10mK.
    \textbf{B}, Same as panel A, but measured at $B_\parallel=165$mT.  Zero resistance renders as bright cyan in a sash between the Sym$_{12}$ and PIP$_2$ states for D$\gtrsim$1. 
    \textbf{C}, $n_{\rm e}$ dependence of $R_{xx}$ measured at fixed $D=1.02V/nm$ and $B_\parallel=165$mT and variable temperatures. 
    Inset: $R(T)$ measured at $n_{\rm e}=-0.57\times10^{12}{\rm cm}^{-2}$ and $D=1.02$V/nm. 
    \textbf{D}, Nonlinear transport in the superconducting state measured at $n_{\rm e}=-0.57\times10^{12}{\rm cm}^{-2}$, $D=1.02$V/nm, and $B_\parallel=165$mT.  
    Inset: $V(I)$ plotted on a log scale.  Dotted line corresponds to $V\propto I^3$, from which we determine $T_{BKT}=26$mK. 
    \textbf{E}, $n_{\rm e}$ dependence of $R_{xx}$ measured at fixed $D=1.02$V/nm and $B_\parallel=165$mT, for variable $B_\perp$.
    \textbf{F}, $B_\perp$-dependent nonlinear transport for fixed   $n_{\rm e}=-0.57\times10^{12}{\rm cm}^{-2}$, $D=1.02$V/nm and $B_\parallel=165$mT. 
    }
    \label{fig:2}
\end{figure*}

The symmetry breaking transitions move to higher $|n_{\rm e}|$ with increasing $|D|$, as expected from a Stoner picture given that increasing $|D|$ enhances the size of the van Hove singularity favoring isospin symmetry breaking states at higher carrier concentration\cite{castro_low-density_2008}. This behavior is analogous to that observed in rhombohedral trilayer graphene (RTG)\cite{zhou_half-_2021}, but the observed phases differ between the two systems. For example, we find no signatures of the annular Fermi sea that hosts superconductivity in RTG, nor do we find a spin-polarized half-metal state.  Finally, the quarter metal state IF$_1$ occupies only a very small domain in the parameter space.  These differences may be tied to subtle  differences in the underlying band wave functions in these two systems.  Further differentiating BBG and RTG, electrical resistivity remains finite at all densities at $B=$0 (Fig. \ref{fig:2}A). 

Most unusually, superconductivity emerges with the application of a finite magnetic field. Fig. \ref{fig:2}B shows the resistivity, measured at a nominal temperature of 10mK and $B_\parallel=165$mT applied in the plane of the sample. A zero-resistance state appears at large $D$ at the apparent transition between the PIP$_2$ and Sym$_{12}$ states (similar data are obtained for $D<0$, see Fig. \ref{fig:s:R_vs_n_minus_p}  in \cite{zhou_supplementary_2021}).  
Figs. \ref{fig:2}C-D show the temperature dependent linear and nonlinear resistivity within this zero-resistance state. 
We define $T_{\rm BKT}=26$mK from the temperature where the voltage $V\propto I^3$. The critical temperature is not found to decrease significantly with applied $B_\parallel$ over the accessible range (see Fig. \ref{fig:s:tc_Bp} in \cite{zhou_supplementary_2021} for additional data). 

In many spin-triplet superconductors non-magnetic impurities play the role of magnetic impurities in conventional superconductors. 
To assess whether such orders are plausible, we estimate the disorder strength, parameterized by the ratio $d=\xi/\ell_{mf}$\cite{abrikosov_contribution_1960} of the superconducting coherence length $\xi$ and the electronic mean free path in the normal state. 
$\xi$ may be estimated from the upper critical field at base temperature of $B_{c\perp}\approx 5$mT (Fig. \ref{fig:2}E-F) from the relation\cite{tinkham_michael_introduction_1975}   $\xi=\sqrt{\Phi_0/(2\pi B_{c\perp})}\approx 250$nm.  This value is comparable to RTG\cite{zhou_superconductivity_2021} and much longer than in moir\'e graphene multilayers\cite{cao_unconventional_2018,cao_pauli-limit_2021}.

\begin{figure}
    \centering
    \includegraphics{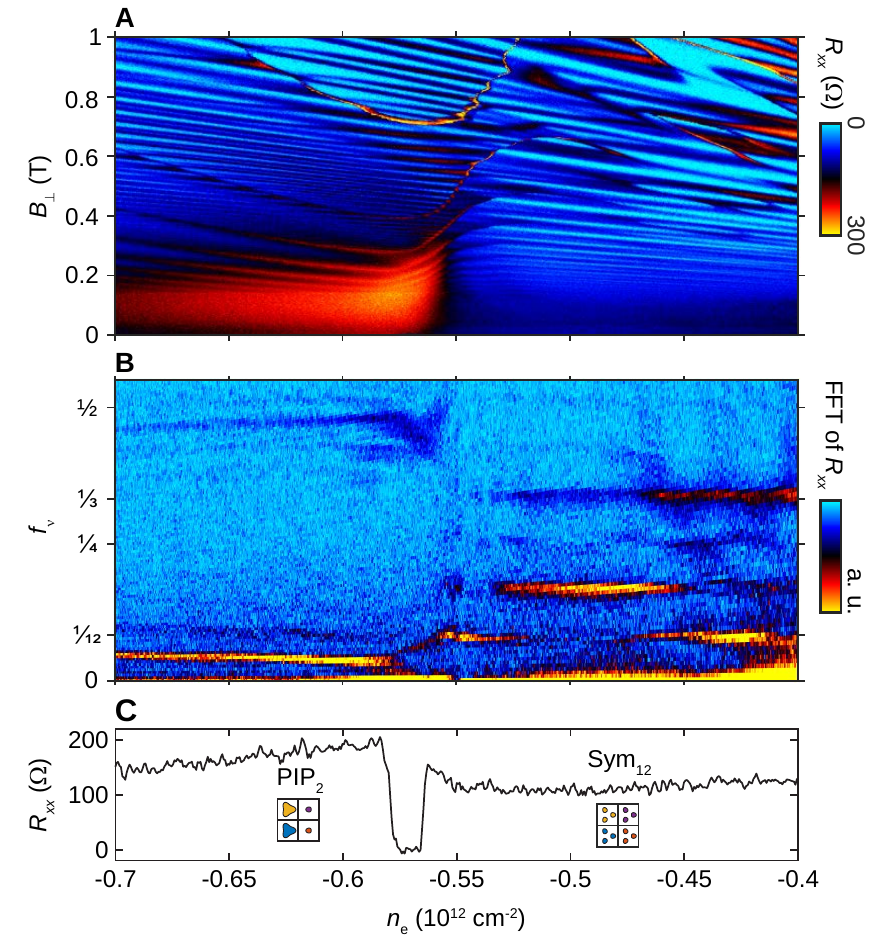}
    \caption{\textbf{Fermiology of the superconducting state.}
    \textbf{A,} $B_\perp$ dependence of $R_{xx}$ at fixed $D=1.02$V/nm and $B_\parallel=165$mT in the $n_{\rm e}$ range near the superconducting phase.
    \textbf{B,} Fourier transform of $R_{xx}(1/B)$, calculated from the data in A. Only data within $0.1$T$<B<0.4$T is used to calculate the result. Data are plotted as a function of $f_\nu$ as in Fig. \ref{fig:1}. 
    \textbf{C,}
    $R_{xx}$ vs $n_{\rm e}$ at $B_\perp=$0, $B_\parallel=165$mT.
    }
    \label{fig:3}
\end{figure}

We may estimate $\ell_{mf}$ from the magnetic field where quantum oscillations are first observed.  This corresponds to $\ell_{mf}\approx2\pi k_f\ell_B^2$, the circumference of a cyclotron orbit\cite{vavilov_magnetotransport_2004}. 
Fig. \ref{fig:3} shows quantum oscillation data in the vicinity of the superconducting state. At $n_{\rm e}\approx -0.57 \times 10^{12} \mathrm{cm}^{-2}$, on the cusp of the superconducting state, two oscillation frequencies are observed. Taking the higher $f_\nu\lessapprox 1/2$, from which we estimate $k_f=\sqrt{2\pi f_\nu |n_{\rm e}|}=0.13 {\rm nm}^{-1}$.  In this regime, the onset field is found to be $\approx$140mT (see Fig. \ref{fig:s:lmf}), giving $\ell_{mf}\approx5\mu$m.  This estimate is comparable to device dimensions\cite{wang_one-dimensional_2013}, so we conclude that $d<0.05$, placing superconductivity deep in the clean limit. Exotic superconducting order parameters are thus not ruled out by disorder considerations. 

BBG and RTG share the same crystal symmetries, differing only in quantitative details.  It is likely then that just as for RTG, the bare fact that superconductivity occurs at a magnetic transition in BBG can be explained by both conventional electron-phonon as well as electronically mediated attraction \cite{vafek_superconductivity_2014,lothman_universal_2017,ojajarvi_competition_2018,ghazaryan_unconventional_2021,chatterjee_inter-valley_2021,dong_superconductivity_2021,cea_superconductivity_2021,szabo_parent_2021,you_kohn-luttinger_2021,chou_acoustic-phonon-mediated_2021}.
We thus turn our attention to new constraints offered by BBG on theories of the superconducting state that seek to capture the qualitative details of the phase diagram.  

In contrast to superconductors in RTG, the fermiology indicates that the superconducting state in BBG emerges from a partially isospin polarized normal state with both majority and minority Fermi surfaces.  
In the domain of the superconducting state, both a high-frequency and a low frequency oscillation are evident (Figs. \ref{fig:3}A-B, and additional data in Fig. \ref{fig:s:sdh_full}), evolving continuously from the peaks in the PIP$_2$ phase. The normal state has a somewhat distinct fermiology from the PIP$_2$ state observed at higher $|n_{\rm e}|$, with the two phases showing contrasting $n_{\rm e}$-dependence of both the low- and high-frequency oscillations. 
In the PIP$_2$ state, $df_\nu/dn_{\rm e}$ is negative for the low $f_\nu$ oscillation and positive for the high $f_\nu$ oscillation, with these trends reversing abruptly at the boundary of the superconducting state. Intriguingly, the low frequency peak in the superconducting regime continuously interpolates between the area of the small Fermi surface in the PIP$_2$ state at high $|n_{\rm e}|$ and the area of a single Fermi pocket of the  Sym$_{12}$ state, suggestive of a continuous transition between the two.  This picture implicitly requires significant reconstruction of the Fermi surfaces relative to the single particle band structure through either inter-valley coherence or the development of nematic order. Alternatively, the superconductor may arise from a partially isospin polarized phase that is distinct from the PIP$_2$ phase, existing between the PIP$_2$ and Sym$_{12}$ phases and breaking a different set of spin, valley, or lattice symmetries.  

\begin{figure*}
\centering
\includegraphics{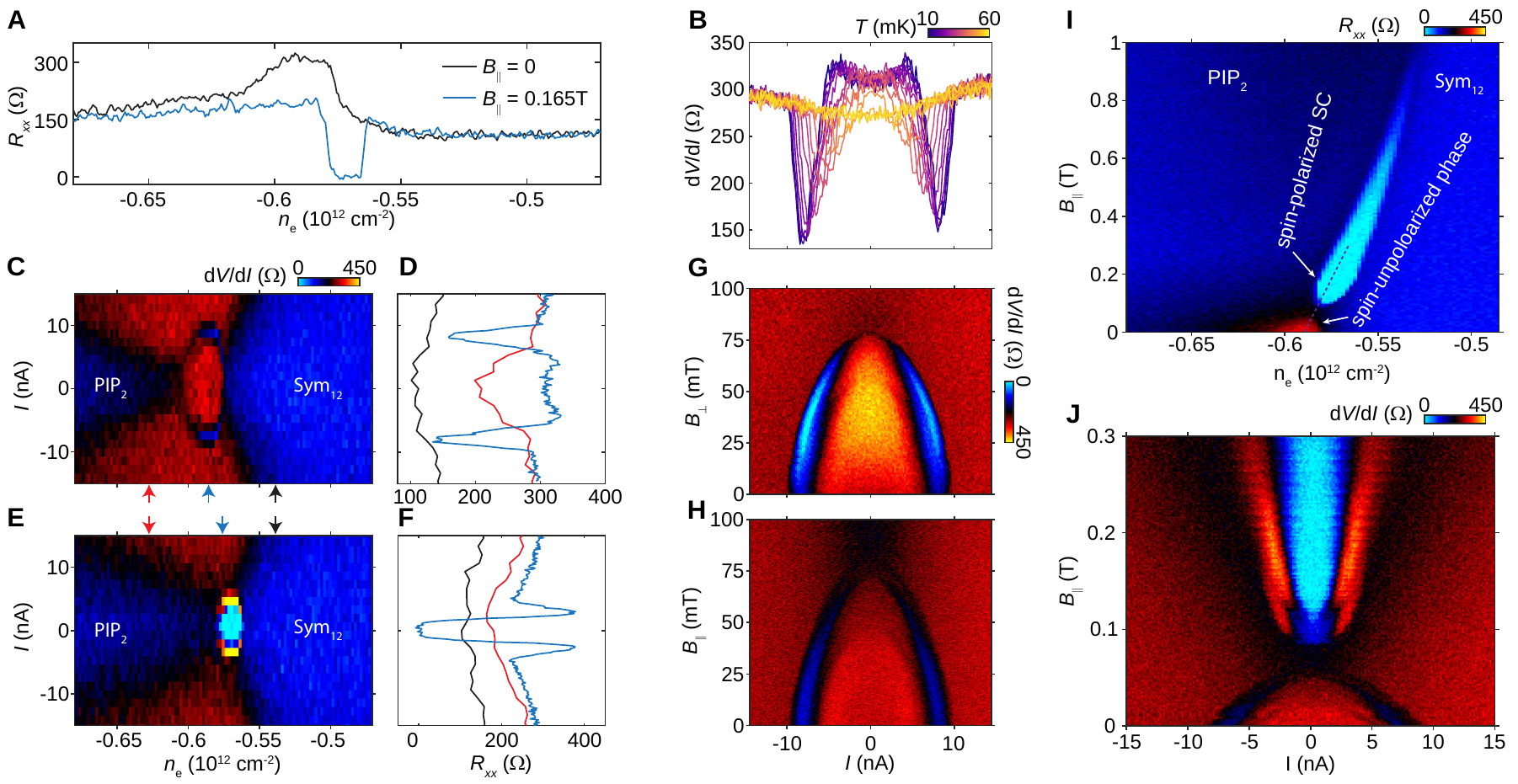}
\caption{{\bf Magnetic-field induced phase transitions.}
\textbf{A,} $n_{\rm e}$-dependent resistivity measured at $D=$1.02V/nm, and $B_\parallel$=0.165T (blue) and zero magnetic field (black).
\textbf{B,} Nonlinear resistivity $n_{\rm e}=-0.59\times10^{12}{\rm cm}^{-2}$, $D=1.02$V/nm, and zero magnetic field at variable $T$. 
\textbf{C,} $n_{\rm e}$-dependent nonlinear $dV/dI$ measured at $D=1.02$V/nm, zero magnetic field
\textbf{D,}  $dV/dI$ as a function of $I$ measured at $n_{\rm e}$ values indicated by arrows in panel C.
\textbf{E,}  Same as panel C with $B_\parallel=$0.165T.    
\textbf{F,}  Same as panel D with $B_\parallel=$0.165T.    
\textbf{G,} $dV/dI$ as a function of $B_\perp$ at $n_{\rm e}=-0.59\times10^{12}{\rm cm}^{-2}$, $D=1.02$V/nm, $B_\parallel$=0.
\textbf{H,} Same as panel G, but as a function of $B_\parallel$ with $B_\perp$=0.
\textbf{I,} $B_\parallel$-dependence of linear response resistivity measured at $D = 1.02$V/nm, $B_\perp$=0.
\textbf{J,} $dV/dI$ measured along the trajectory shown by the dashed line in panel I.
}\label{fig:4}
\end{figure*}

Comparing transport measurements at zero and finite $B_\parallel$ (Fig. \ref{fig:4}A) provides additional information. In the absence of a field, the Sym$_{12}$ and PIP$_2$ states are separated by a resistance maximum (Fig. \ref{fig:4}A), where the resistivity shows strong nonlinearity.  Specifically, the resistance is constant up to a sharply defined threshold in the applied current, where it abruptly changes to a reduced value (Fig. \ref{fig:4}B and \ref{fig:sampletwo}). This behavior is reminiscent of phenomena observed  in  charge density wave compounds associated with electric field induced depinning\cite{fleming_sliding-mode_1979}. Both resistance peak and nonlinearity are low-temperature phenomena, appearing only below $T\approx 50$mK as shown in Figs. \ref{fig:4}B and \ref{fig:s:R_vs_T_large_range}.  
Like superconductivity at finite $B_\parallel$ (Fig. \ref{fig:4}E-F), at B=0 the threshold current reaches a maximum in between the Sym$_{12}$ and PIP$_2$ states (Fig. \ref{fig:4}C-D). 
These facts suggest that the nonlinearity is the signature that the low-temperature ground state at zero magnetic field has broken symmetries distinct from the neighboring PIP$_2$ phase.

The threshold observed at B=0 is continuously suppressed by applied magnetic fields (Figs. \ref{fig:4}g-h) independent of the field direction, and  disappears completely for $|B|>75$mT (Fig. \ref{fig:s:R_vs_I_theta_B}).  This suggests that the zero field phase is spin unpolarized, and that the suppression of the nonlinearity is driven by a spin polarization transition. 
As shown in Fig. \ref{fig:4}I, the boundary between the PIP$_2$ and Sym$_{12}$ phases shifts to lower $|n_{\rm e}|$ with increasing $B_\parallel$. Tracing the nonlinear transport along the boundary between the Sym$_{12}$ and PIP$_2$ phases (Fig. \ref{fig:4}J) reveals that the suppression of the resistive state coincides precisely with the onset of superconductivity. We conclude that the observed superconductivity arises as soon as the Zeeman energy is sufficient to spin polarize the electron system, destroying the B=0 phase and turning on the superconducting ground state.  

The current results introduce significant new constraints to any universal theory of superconductivity in graphene systems---assuming such a theory exists.  
In particular, the difference in Fermi surface topology between the BBG, RTG, and moir\'e systems in their respective superconducting regimes suggests that Fermi surface details are not central to the superconducting mechanism.  In contrast, proximity to an isospin ordered phase is a generic feature of both moir\'e and crystalline graphene superconductors, suggestive of a fluctuation mediated or other purely repulsive mechanism. 

However, we note that our experiments do not yet rule out a phonon mediated mechanism, where generic pairing only leads to observable superconducting $T_C$ in a narrow density range and for a specific underlying isospin ordered phase.  
With respect to resolving the mechanism of superconductivity, the greatest impact of the current work to be practical: the stability of BBG allows exceptionally high-quality systems to be made with high yield and reproducibility.  Moreover, van Hove singularities of the type explored here and in RTG are generic to all graphene multilayers, so we expect field-effect controlled superconductivity to be a widespread phenomenon in graphene allotropes with sufficiently low disorder.

\section*{acknowledgments}
The authors acknowledge discussions with E. Berg, S. Das Sarma, Y.-Z. Chou, A. Ghazaryan, L. Levitov, A.  Macdonald, M. Serbyn, C. Varma, and  M. Zaletel.  
This project was primarily funded by the Department of Energy under DE-SC0020043.  
Support to purchase the cryogen-free dilution refrigerator  was provided by Army Research Office under award W911NF-17-1-0323.
 AFY acknowledges the support of the Gordon and Betty Moore Foundation under award GBMF9471 and the Packard Foundation
under award 2016-65145 for general group activities.  
KW and TT acknowledge support from the Elemental Strategy Initiative conducted by the MEXT, Japan (Grant Number JPMXP0112101001) and  JSPS KAKENHI (Grant Numbers 19H05790, 20H00354 and 21H05233).

\section*{data and materials availability}
All data shown in the main text and supplementary materials are available in the Dryad data repository\cite{original_data}.

\clearpage
\newpage
\pagebreak

\onecolumngrid

\begin{center}
\textbf{\large Supplementary information for ``Isospin magnetism and spin-triplet superconductivity in Bernal bilayer graphene'' }\\[5pt]

\end{center}
\setcounter{equation}{0}
\setcounter{figure}{0}
\setcounter{table}{0}
\setcounter{page}{1}
\setcounter{section}{0}
\makeatletter
\renewcommand{\theequation}{S\arabic{equation}}
\renewcommand{\thefigure}{S\arabic{figure}}
\renewcommand{\thepage}{S\arabic{page}}

\textbf{This PDF file includes:}
\begin{itemize}
    \item Materials and Methods
    \item Supplementary figures
\end{itemize}

\section{Materials and Methods}
The bilayer graphene, few-layer graphite and hBN flakes are prepared by mechanical exfoliation of bulk crystals. The van der Waals heterostructures are assembled using a dry transfer technique\cite{wang_one-dimensional_2013} with poly-propylene carbonate (PPC) film for Device A and poly bisphenol a carbonate (PC) film for Device B. Device  A, shown in the main text, was previously studied in references \cite{zibrov_tunable_2017} and \cite{island_spinorbit-driven_2019}, where it is denoted Device C and C1, respectively. These references also describe the fabrication methods in more detail. 
To form graphite contacts for device B, the graphite flake was patterned with anodic oxidation using an atomic force microscope\cite{li_electrode-free_2018}. Both devices have a dual-graphite gate structure\cite{zibrov_tunable_2017} to minimize the charge disorder and to allow independent control of the carrier electron density ($n_{\rm e}$) and the electrical displacement field ($D$). 

All electrical measurements are performed in a dilution refrigerator equipped with a 9T/1T/1T superconducting vector magnet. The vector-field control is essential for $B_\parallel$-dependence measurements, allowing precise control of the field direction, in particular removing residual $B_\perp$. Transport measurements are performed using lock-in techniques at a frequency $f< 45$Hz to reduce the electronic noise. Low-pass electronic filtering basd on \cite{kuemmeth_reducing_2015} is applied to lower the electron temperature. Penetration field capacitance was measured using a capacitance bridge circuit with an FHX15X high electron mobility transistor serving as an in-situ impedance transformer, as described in \cite{zibrov_tunable_2017}. An excitation frequency of 54245.12Hz was used to obtain the capacitance data. To convert the measured values to inverse electronic compressibility $\kappa$ we use low-magnetic field Landau levels as a calibration for perfect screening and perfect penetration, a procedure described in detail in Ref. \cite{zhou_half-_2021}.

\section{Supplementary Figures}
\begin{figure*}[h]
    \centering
    \includegraphics{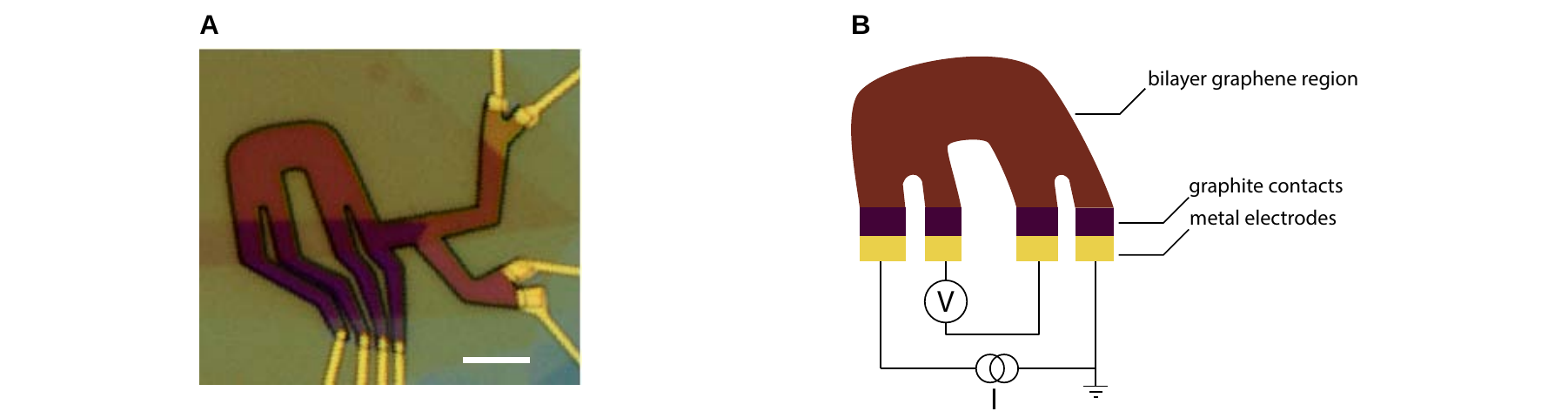}
    \caption{\textbf{Geometry of Device A and transport measurement wiring.}
    \textbf{A,} Optical micrograph of the sample studied. Scale bar represents 5$\mu$m.  The entire sample consists of a graphite-hBN-BBG-hBN-graphite stack, with the graphite layers functioning as gates and hexagonal boron nitride layers functioning as gate dielectrics.  Additional graphite serves as contacts to the BBG, as shown in the figure.  Both top- and bottong hBN flakes are 38nm thick. 
    \textbf{B,} Configuration of the transport measurements.}
    \label{fig:s:sample}
\end{figure*}

\begin{figure*}[h]
    \centering
    \includegraphics{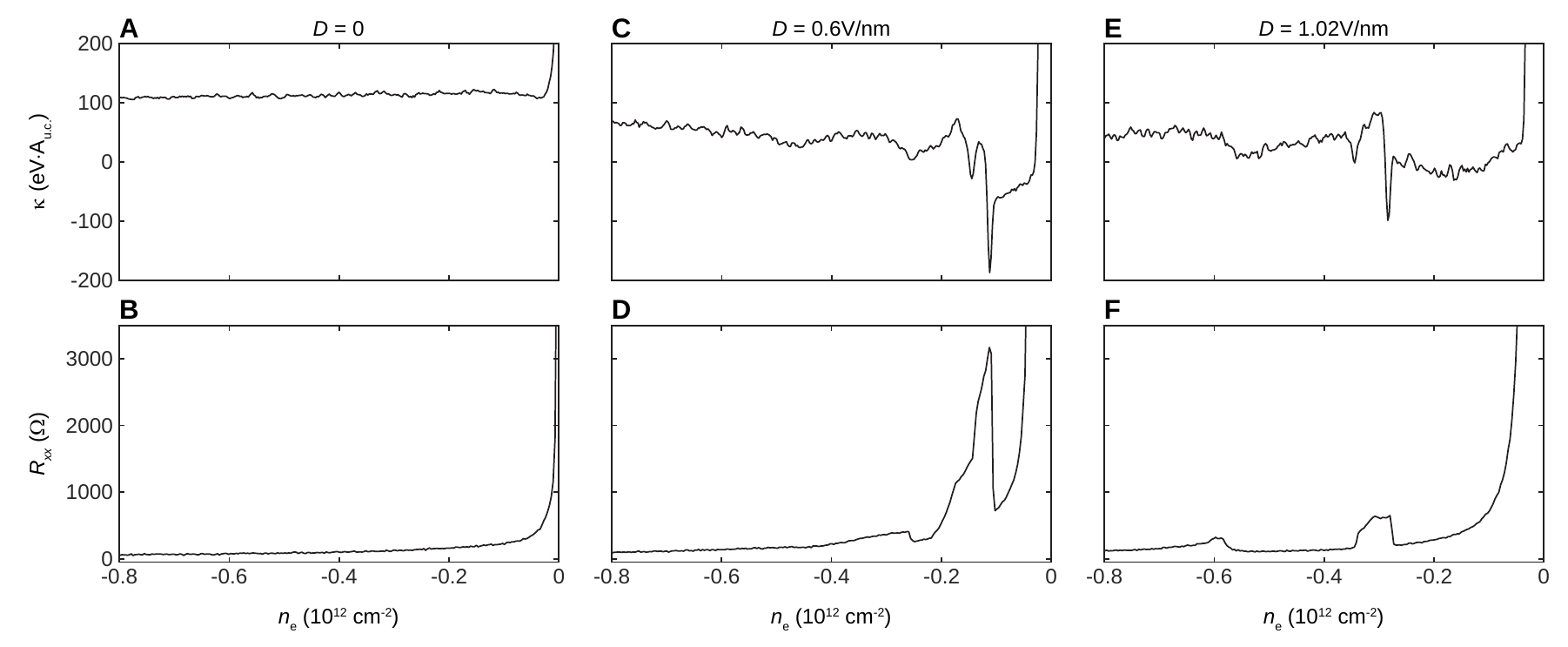}
    \caption{\textbf{Comparison of resistivity and inverse compressibility at zero magnetic field} 
    \textbf{A,} Resistivity versus carrier density measured at $D=0$.
    \textbf{B,} Inverse compressibility versus carrier density measured at $D=0$.
    \textbf{C,} Same as A, measured at $D=0.6$V/nm.
    \textbf{D,} Same as B, measured at $D=0.6$V/nm.
    \textbf{E}, Same as A and C, measured at $D=1.02$V/nm.
    \textbf{F}, Same as B and D, measured at $D=1.02$V/nm.
    }
    \label{fig:s:r_cp_linecuts}
\end{figure*}

\begin{figure*}[h]
    \centering
    \includegraphics{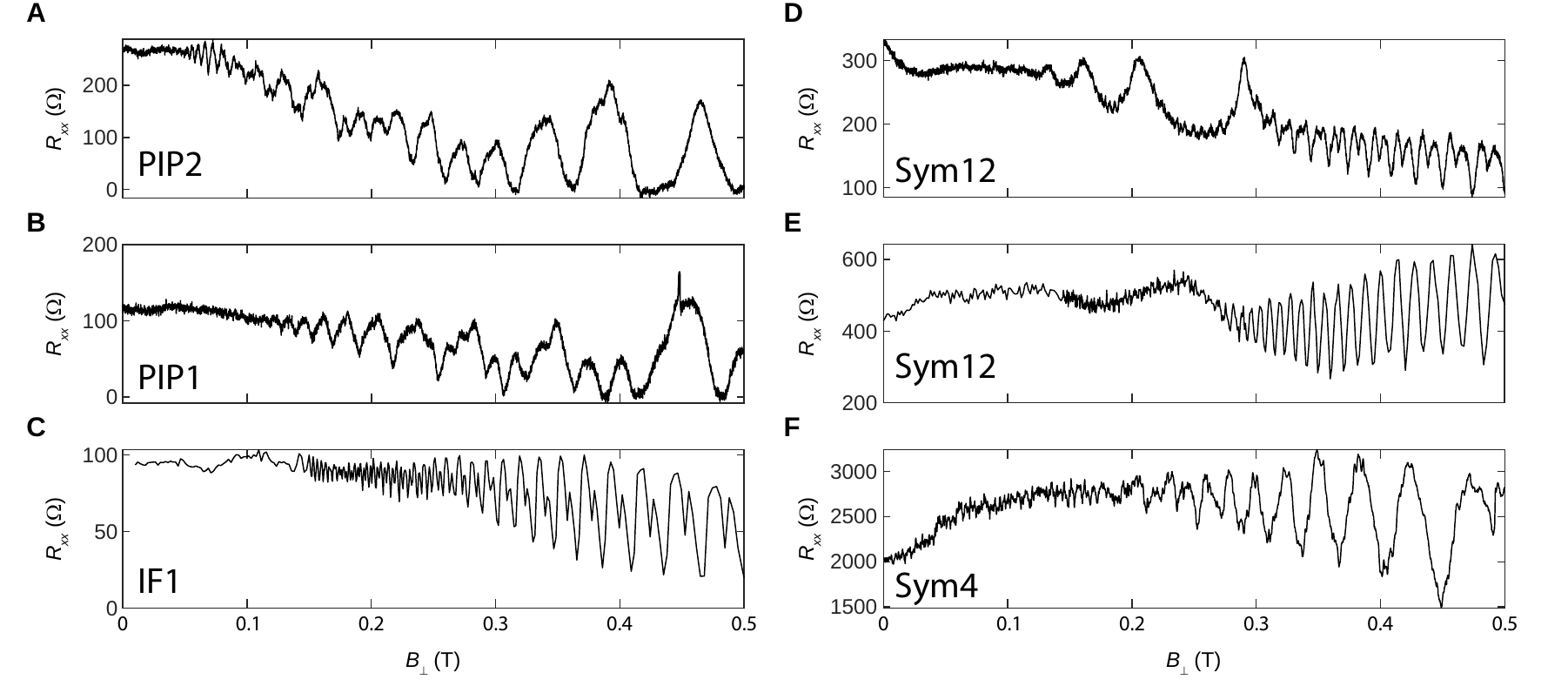}
    \caption{
    \textbf{Quantum oscillation data used to generate Fig \ref{fig:1}f. }
    Resistivity as a function of $B_\perp$ at fixed $n_{\rm e}$ and $D$. 
    Panels A to E correspond to traces used to generate Fig. \ref{fig:1}F from bottom to top respectively.
    }
    \label{fig:s:original_osc}
\end{figure*}

\begin{figure*}[h]
    \centering
    \includegraphics{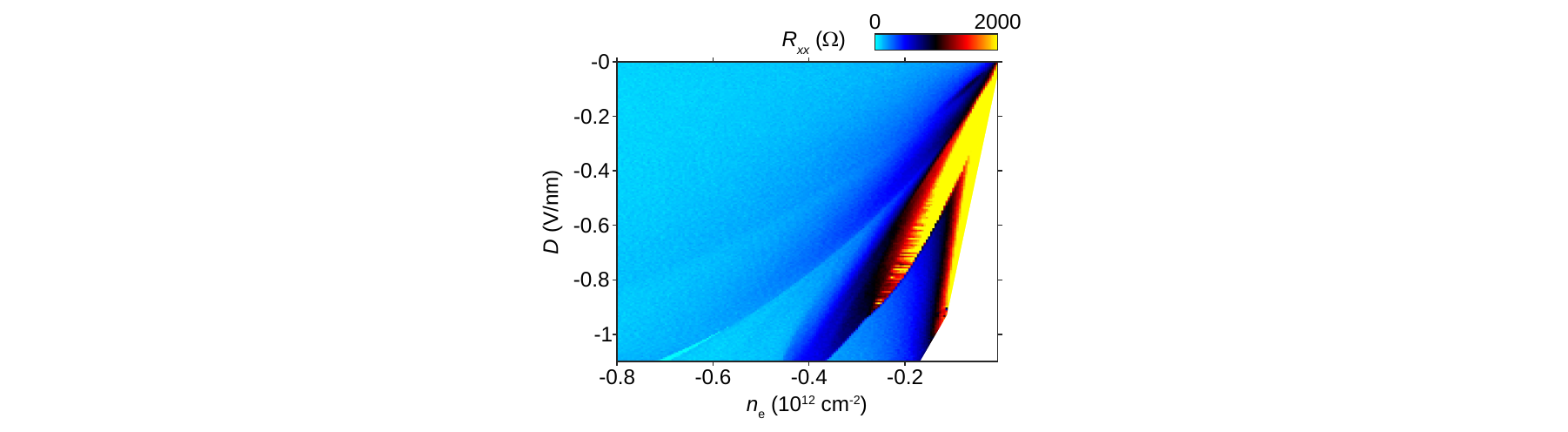}
    \caption{\textbf{Superconducting phase at negative displacement field.} Measurement was performed at $B_{\parallel}=$0.165T. Superconductivity is visible as a bright cyan sash between $n_{\rm e}\approx -.6$ and $n_{\rm e}\approx -.7\times 10^{12}$cm$^{-2}$ and $|D|\gtrsim 1$ V/nm.}
     \label{fig:s:R_vs_n_minus_p}
\end{figure*}

\begin{figure*}[h]
    \centering
    \includegraphics{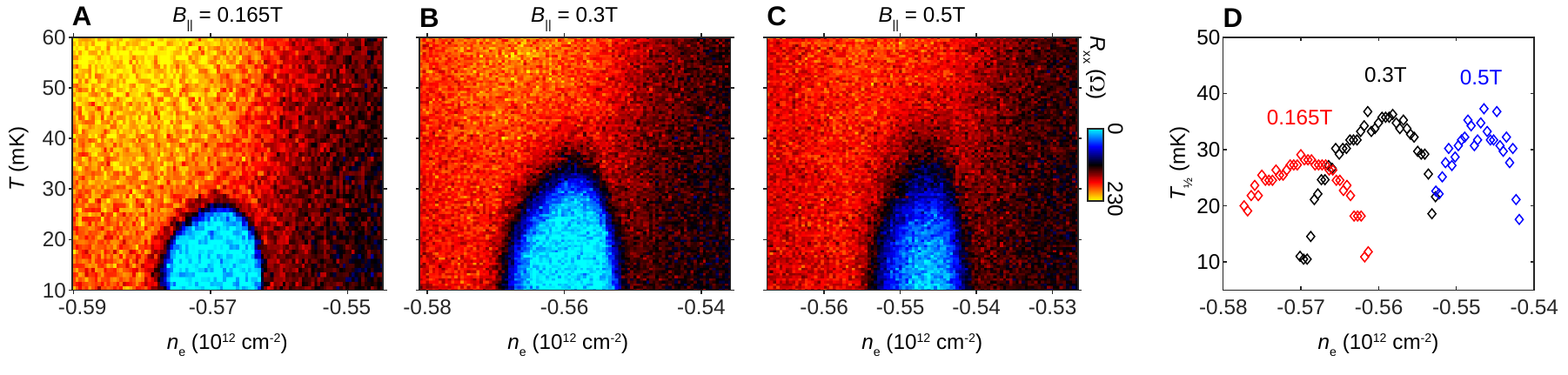}
    \caption{\textbf{$B_\parallel$-dependence of the superconducting phase.}
    \textbf{A,} $R_{xx}$ versus carrier density and temperature measured at $D=1.02$V/nm and $B_\parallel=0.165$T.
    \textbf{B,} Same as A, measured at $B_\parallel=0.3$T.
    \textbf{C,} Same as A and B, measured at $B_\parallel=0.5$T.
    \textbf{D,} $T_{1/2}$ versus carrier density extracted from the data in panel A-C. $T_{1/2}$ is the temperature at which the resistance is 50\% of the normal state resistance.}
    \label{fig:s:tc_Bp}
\end{figure*}

\begin{figure*}[h]
    \centering
    \includegraphics{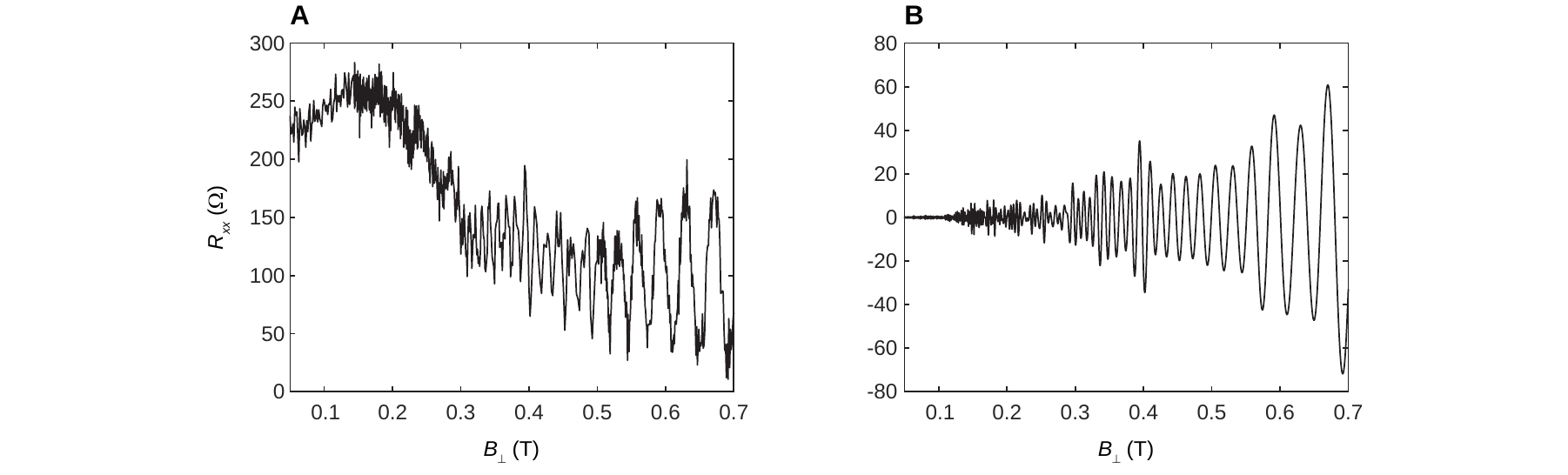}
    \caption{\textbf{Onset $B_\perp$ of quantum oscillations oscillation.} 
    \textbf{A,} Measured $R_{xx}$ as a function of $B_\perp$ at $n_{\rm e}=-0.57\times10^{12}{\rm cm}^{-2}$, $D=1.02$V/nm near the superconducting state.  
    \textbf{B,} The same data as in panel A, after applying a Fourier domain band bass filter to isolate the higher frequency oscillation.  Oscillations are visible well below 200mT.}  
    \label{fig:s:lmf}
\end{figure*}

\begin{figure*}[h]
    \centering
    \includegraphics{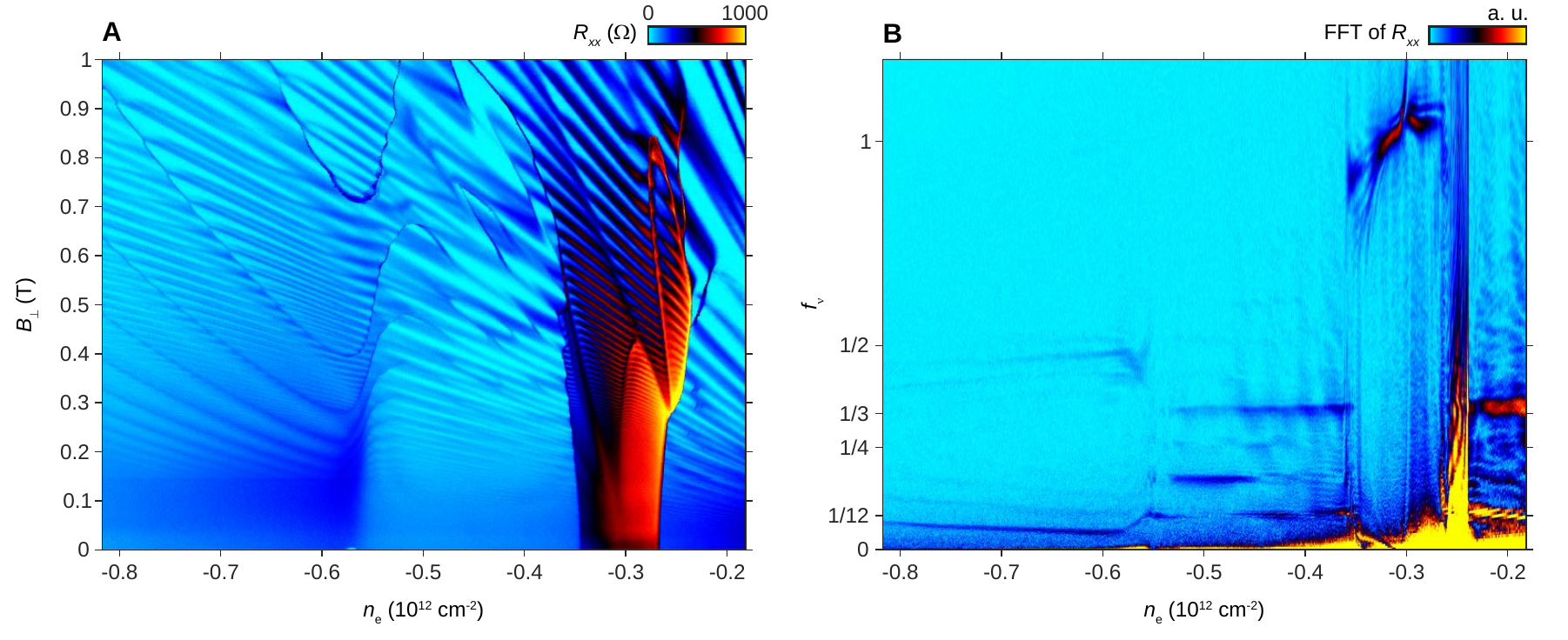}
    \caption{\textbf{Large density range fermiology} 
    \textbf{A,} $B_\perp$ and $n_{\rm e}$ dependent resistivity measured at $D=1.02$V/nm, showing a larger range of density. 
    \textbf{B,} Fourier transform of $R_{xx}$ shown in panel A over the range $0<B_\perp<0.4$T, as described in the main text. }
    \label{fig:s:sdh_full}
\end{figure*}

\begin{figure*}[h]
    \centering
    \includegraphics{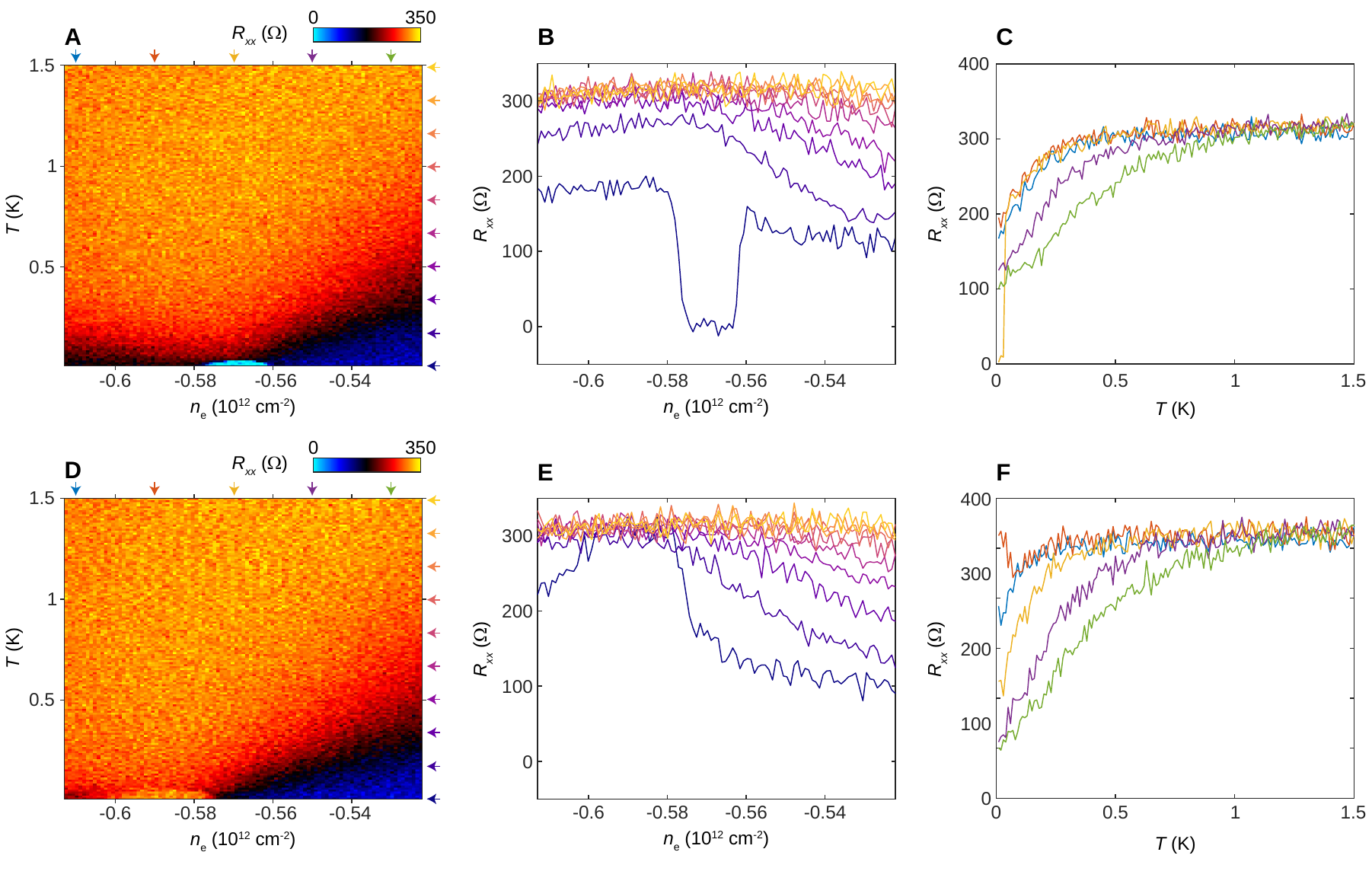}
    \caption{\textbf{Temperature dependent resistance near the SC and spin-unpolarized phase.}
    \textbf{A,} Resistivity versus carrier density and temperature measured at $D=1.02$V/nm and $B_\parallel=$0.165T.
    \textbf{B,} Linecuts of data in panel A at fixed temperature indicated by colored arrows in panel a.
    \textbf{C,} Linecuts of data in panel A at fixed carrier density indicated by colored arrows in panel A.
    \textbf{D}, Same as A, measured at $B_\parallel=$0.
    \textbf{E}, Same as B, measured at $B_\parallel=$0.
    \textbf{F}, Same as C, measured at $B_\parallel=$0.
    }
    \label{fig:s:R_vs_T_large_range}
\end{figure*}

\begin{figure*}[h]
    \centering
    \includegraphics{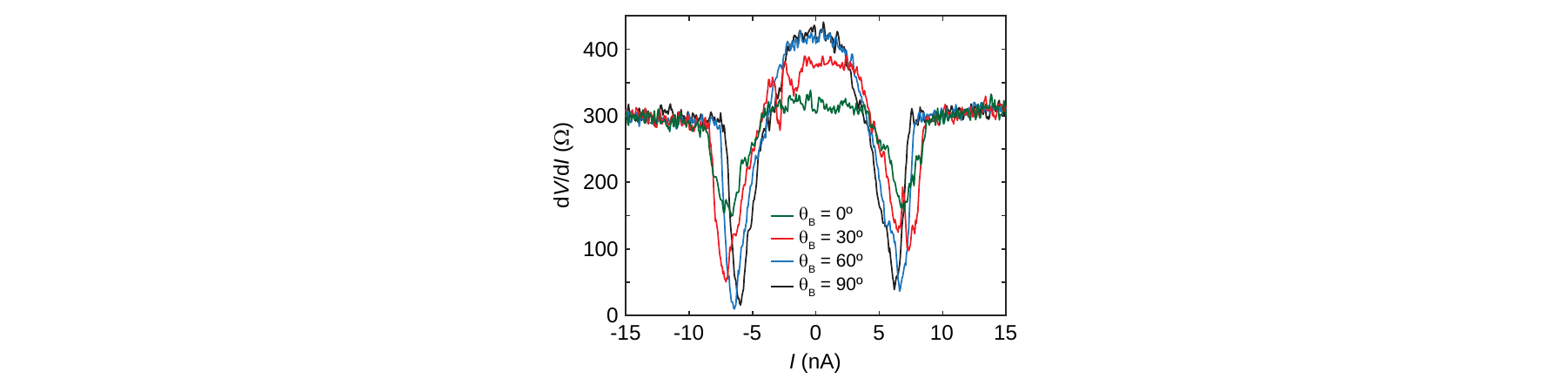}
    \caption{\textbf{$\mathbf{\theta_B}$-dependence of the nonlinear transport in the spin-unpolarized phase.} $dV/dI$ measured at $n_{\rm e}=-0.57\times10^{12}{\rm cm}^{-2}$, $D=1.02$V/nm with a 50mT magnetic field $B$ applied along the direction defined by $\theta_B$, where $B_\perp=B\cos{\theta_B}$, $B_\parallel=B\sin{\theta_B}$.}
   \label{fig:s:R_vs_I_theta_B}
\end{figure*}

\begin{figure*}[h]
    \centering
    \includegraphics{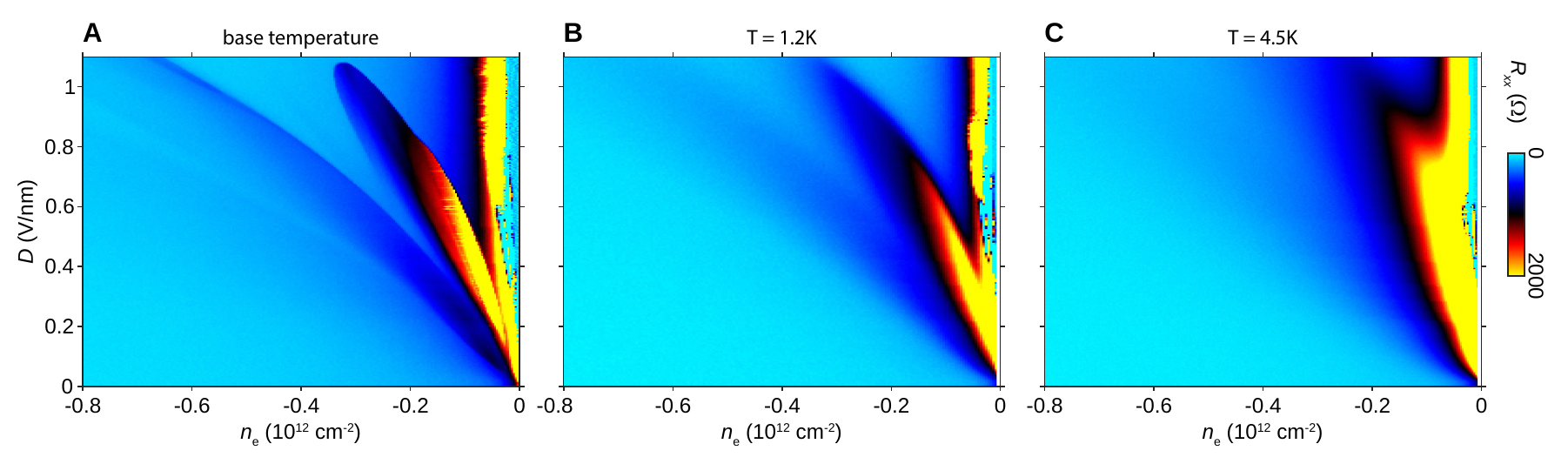}
    \caption{\textbf{Temperature dependence of the isospin phase diagram at zero magnetic field.}
    \textbf{A,} $R$ vs $n_{\rm e}$ and $D$ measure at base temperature.
    \textbf{B,} Same as panel A, measured at $T=$1.2K.
    \textbf{C,} Same as panel A and B, measured at $T=$4.5K.}
    \label{fig:s:R_vs_np_T}
\end{figure*}

\begin{figure*}[h]
    \centering
    \includegraphics{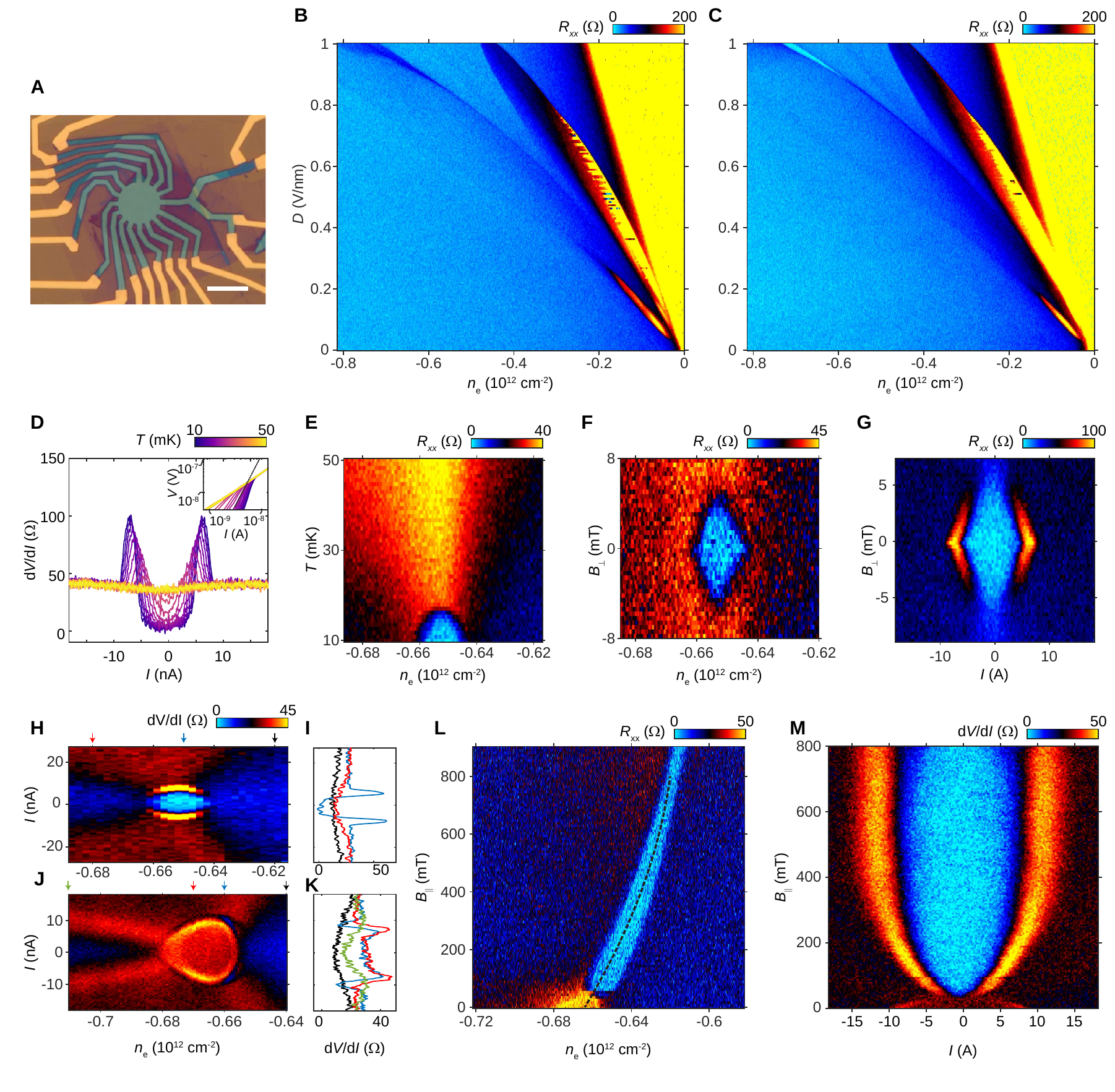}
    \caption{\textbf{Data from Device B.}
    \textbf{A,} Optical micrograph of the sample. Scale bar represents 10$\mu$m.  
    For this device, the gate dielectrics were 9nm and 18nm thick.  
    \textbf{B,} $R_{xx}$ measured at fixed $B_\parallel=0$ at a nominal temperature of 10mK.
    \textbf{C,} Same as panel B, but measured at $B_\parallel=150$mT.
    \textbf{D}, Nonlinear transport in the superconducting state measured at $n_{\rm e}=-0.652\times10^{12}{\rm cm}^{-2}$, $D=0.96$V/nm, and $B_\parallel=150$mT. Inset: $V(I)$ plotted on a log scale.  Dotted line corresponds to $V\propto I^3$, from which we determine $T_{BKT}=22$mK.
    \textbf{E}, $n_{\rm e}$ and $T$ dependence of $R_{xx}$ measured at fixed $D=0.96$V/nm and $B_\parallel=150$mT.
    \textbf{F}, $n_{\rm e}$ and $B_\perp$ dependence of $R_{xx}$ measured at fixed $D=0.96$V/nm and $B_\parallel=150$mT.
    \textbf{G,} $B_\perp$-dependent nonlinear transport for fixed   $n_{\rm e}=-0.652\times10^{12}{\rm cm}^{-2}$ and $D=0.96$V/nm. 
    \textbf{H,} $n_{\rm e}$-dependent nonlinear $dV/dI$ measured at $D=0.96$V/nm, $B_\parallel=150$mT. 
    \textbf{I,}  $dV/dI$ as a function of $I$ measured at the same $D$ and $B_\parallel$ as those in panel h and $n_{\rm e}$ values indicated by arrows in panel H.
    \textbf{J,} Same as panel h, $B_\parallel=0$.
    \textbf{K,}  $dV/dI$ as a function of $I$ measured at the same $D$ and $B_\parallel$ as those in panel j and $n_{\rm e}$ values indicated by arrows in panel J.
    \textbf{L,} $B_\parallel$-dependence of linear response resistivity measured at $D = 0.96$V/nm, $B_\perp$=0.
    \textbf{M,} $dV/dI$ measured along the trajectory shown by the dashed line in panel L.
    }\label{fig:sampletwo}
\end{figure*}
\end{document}